%% file: main.tex
\documentclass[sigconf]{acmart}
\acmConference[EASE 2026]{The 30th International Conference on Evaluation and Assessment in Software Engineering}{9–12 June, 2026}{Glasgow, Scotland, United Kingdom}
\usepackage{graphicx} 
\usepackage{hyperref}
\usepackage{amsmath}
\usepackage{wrapfig}
\usepackage{listings}
\usepackage{xcolor}
\usepackage{multirow}
\usepackage{tcolorbox}
\usepackage{makecell}
\usepackage{tikz}
\usepackage{xcolor,colortbl}
\lstdefinestyle{console}{
    basicstyle=\ttfamily\footnotesize,
    backgroundcolor=\color{gray!10},
    frame=single,
    breaklines=true,
    postbreak=\mbox{\textcolor{red}{$\hookrightarrow$}\space},
}
\setcitestyle{maxbibnames=3}
\usepackage{booktabs}
\usepackage{colortbl}
\usepackage[table]{xcolor}
\definecolor{corrTop}{RGB}{26,152,80}     
\definecolor{corrMid}{RGB}{166,217,106}   
\definecolor{corrLow}{RGB}{254,224,139}   

\definecolor{promptbg}{RGB}{245,245,245}      
\definecolor{promptborder}{RGB}{180,180,180}  
\definecolor{prompttext}{RGB}{0,0,30}       
\definecolor{promptbreak}{RGB}{0,102,204}     

\lstdefinestyle{prompt}{
    backgroundcolor=\color{promptbg},
    frame=single,
    rulecolor=\color{promptborder},
    basicstyle=\ttfamily\scriptsize\color{prompttext}, 
    breaklines=true,
    columns=fullflexible,
    framerule=0.4pt,             
    xleftmargin=0.8em,           
    xrightmargin=0.8em,
    aboveskip=0.3em,             
    belowskip=0.3em,
    postbreak=\mbox{\textcolor{promptbreak}{$\hookrightarrow$}\space}, 
    captionpos=b,
    keepspaces=true,             
    showstringspaces=false       
}

\lstdefinestyle{code}{
    language=Python,
    basicstyle=\ttfamily\footnotesize,          
    keywordstyle=\color{blue}\bfseries,  
    commentstyle=\color{green!60!black}\itshape, 
    stringstyle=\color{orange},          
    showstringspaces=false,              
    breaklines=true,                     
    breakatwhitespace=true,              
    frame=single,                        
    framerule=0.5pt,                     
    rulecolor=\color{gray!70},           
    numbers=left,                        
    numberstyle=\tiny\color{gray},       
    stepnumber=1,                        
    tabsize=4,                            
    captionpos=b,                         
    xleftmargin=5pt, xrightmargin=5pt,   
    backgroundcolor=\color{gray!10}       
}

\usepackage{multirow}
\usepackage{tabularx}
\usepackage{subcaption}
\usepackage{graphicx}
\usepackage{tcolorbox}
\usepackage{amsmath}
\usepackage{amsfonts}
\usepackage{fancyvrb}
\usepackage{xcolor}
\usepackage{wrapfig}

\newtcolorbox{summarybox}{
  boxrule=0.3mm,  
  left=0.5mm,      
  right=0.5mm,     
  top=0.5mm,       
  bottom=0.5mm,    
}

\newcounter{observation}
\renewcommand{\theobservation}{O\arabic{observation}}

\newtcolorbox{observation}[2][]{%
    colback=yellow!10!white,    
    colframe=black,             
    rounded corners,
    arc=5pt,
    boxrule=1pt,
    left=5pt,
    right=5pt,
    top=1pt,
    bottom=1pt,
    before skip=7pt,
    after skip=7pt,
    #1,                         
    before upper={%
        \refstepcounter{observation}
        \phantomsection
        \ifx&#2&\else
            \label{#2}
        \fi
        \noindent\textbf{Observation \theobservation: }%
    }
}

\newtcolorbox{takeaway}[1][]{
    colback=blue!10!white,    
    colframe=black,             
    rounded corners,            
    arc=5pt,                   
    boxrule=1pt,               
    left=5pt,                 
    right=5pt,                
    top=1pt,                   
    bottom=1pt, 
    before skip=7pt, 
    after skip=7pt,  
    label={#1}                         
}

\newcounter{k}
\newcounter{rec}
\title{Evaluating the Environmental Impact of using SLMs and Prompt Engineering for Code Generation}
\author{Md Afif Al Mamun}
\email{afif.mamun@ucalgary.ca}
\affiliation{
    \institution{University of Calgary} 
    \country{Canada}
 } 

\author{Sayan Nath}
\email{sayan.nath@ucalgary.ca}
\affiliation{
    \institution{University of Calgary} 
    \country{Canada}
 } 

\author{Gias Uddin}
\email{guddin@yorku.ca}
\affiliation{
    \institution{York University} 
    \country{Canada}
 } 

\author{Novarun Deb}
\email{novarun.deb@ucalgary.ca}
\affiliation{
    \institution{University of Calgary} 
    \country{Canada}
 } 

\input{abstract}

\begin{document}


\maketitle

\input{introduction_new}
\input{background}

\input{experimental-setup_new}

\input{results_new}
\input{discussion}
\input{conclusion}
\bibliographystyle{abbrv}
\bibliography{references}

\end{document}

%% file: abstract.tex
\begin{abstract}

The shift from cloud-hosted Large Language Models (LLMs) to locally deployed open-source Small Language Models (SLMs) has democratized AI-assisted coding; however, it has also decentralized the environmental footprint of AI. While prompting strategies—such as Chain-of-Thought and ReAct—serve as external mechanisms for optimizing code generation without modifying model parameters, their impact on energy consumption and carbon emissions remains largely invisible to developers. This paper presents the first systematic empirical study investigating how different prompt engineering strategies in SLM-based code generation impact code generation accuracy alongside sustainability factors. We evaluate six prominent prompting strategies across 11 open-source models (ranging from 1B to 34B parameters) using the HumanEval+ and MBPP+ benchmarks. By measuring Pass@1 accuracy alongside energy (kWh), carbon emissions (kgCO$_2$eq), and inference latency, we reveal that sustainability often decouples from accuracy, allowing significant environmental optimizations without sacrificing performance. Our findings indicate that Chain-of-Thought, being a simpler prompting technique, can provide a near-optimal balance between reasoning capability and energy efficiency. Conversely, multi-sampling strategies often incur disproportionate costs for marginal gains. Finally, we identify grid carbon intensity as the dominant factor in deployment-time emissions, highlighting the need for practitioners to consider regional energy profiles. This work provides a quantitative foundation for "green" prompt engineering, enabling developers to align high-performance code generation with ecological responsibility.

\textbf{Replication Package.} \url{https://anonymous.4open.science/r/cg-sustainability-B784}
\end{abstract}

%% file: introduction_new.tex
\section{Introduction}
The rapid adoption of large language models (LLMs) for code generation has transformed modern software engineering, enabling automated code generation, bug fixing, and test generation at scale \cite{chen2021evaluating, liu2024exploring, meng2024empirical}. Tools such as GitHub Copilot \footnote{\url{https://github.com/features/copilot}} and ChatGPT \footnote{\url{https://chatgpt.com/}} now support developers across diverse programming tasks, significantly accelerating development workflows and reducing routine effort while LLM-powered IDEs \cite{wang2024openhands, dong2025survey} are getting traction. The landscape of code generation has undergone a fundamental transformation with the emergence of open-source small language models (SLMs) that can often be used locally by end users. Unlike proprietary tools requiring cloud-based subscriptions, these open-source alternatives provide accessibility, data privacy, and independence from vendor-controlled inference pipelines. However, this democratization introduces a critical sustainability challenge: end-user deployment places the full environmental burden of inference directly on developers' hardware and regional electricity consumption~\cite{alibaba2026carbon}.

While prior work has quantified the energy costs of inferences based on cloud infrastructures, which accounts for approximately 60\% of total AI lifecycle emissions~\cite{columbia2023ai}, the sustainability implications of decentralized open-source code generation remain systematically unexplored. While cloud-based inference outsources both computation and environmental responsibility to centralized data centers, local deployment places the burden of energy efficiency directly on developers' hardware and electricity consumption. Individual developers running inference on personal or organizational machines incur the full environmental cost of their inference workload---potentially across thousands of development interactions per week. When aggregated across millions of developers globally adopting local open-source code generation, these seemingly modest per-query emissions compound into substantial cumulative impact, yet remain almost entirely invisible to practitioners. This gap is consequential because local deployment differs fundamentally from centralized systems: grid carbon intensity varies by regions, and developers typically lack visibility into the energy and carbon implications of their model and prompting choices. Without systematic evidence, practitioners adopting accuracy-only optimization inadvertently incur unnecessary environmental costs.

Concurrently, prompt engineering has become a primary mechanism for improving code generation without retraining models. Techniques such as Chain of Thought (CoT)~\cite{wei2022chain} and ReAct~\cite{yao2022react} have demonstrated substantial improvements in reasoning and code-generation accuracy~\cite{li2025structured}. For developers using open-source models, the choice of prompting strategy is often the most practical lever for operational control, as it requires no additional hardware or infrastructure. Importantly, different prompting strategies can influence the length, structure, and complexity of model outputs, which directly affects the computational and energy cost of each inference. Despite this, existing evaluations overwhelmingly prioritize accuracy, providing minimal insight into the environmental impact of prompt design. This gap is particularly significant for code generation, where high-frequency interactions can rapidly accumulate inference costs.

To the best of our knowledge, no prior work has systematically quantified the sustainability trade-offs of prompting strategies in code generation using SLMs. This paper presents the first comprehensive empirical study examining the impact of prompting approaches on both accuracy and environmental impact. We evaluate six prompting strategies---(1) Direct, (2) Chain-of-Thought \cite{wei2022chain}, (3) Program-of-Thought \cite{chen2022program}, (4) Self-Consistency \cite{wang2022self}, (5) Least-to-Most \cite{zhou2022least}, and (6) ReAct \cite{yao2022react}---across 11 open-source models (1B--34B parameters) on HumanEval+ \cite{chen2021evaluating} and MBPP+ \cite{austin2021program} benchmarks, measuring Pass@1 accuracy alongside energy consumption (kWh), carbon emissions (kg CO$_2$eq), inference latency, and token utilization across heterogeneous hardware and geographic regions. To ensure a realistic evaluation, all strategies are implemented within an established prompting framework \cite{khattab2023dspy} that requires canonical, machine-parsable code outputs—reflecting practical developer workflows.

We find that code generation sustainability decouples from accuracy, enabling environmental improvements without accuracy sacrifice. Chain-of-Thought balances near-optimal accuracy with substantially lower overhead than complex reasoning frameworks, while multi-sampling strategies incur disproportionate token costs without proportional accuracy gains when used with SLMs. We also find that inference time predicts emissions far more strongly than token count alone, while grid carbon intensity is the dominant determinant of deployment-time emissions.

%% file: background.tex
\section{Background \& Related Work}
\label{sec:background}

This section reviews prior work relevant to our study, focusing on three complementary research directions: the use of small language models for code generation, advances in prompt engineering for improving reasoning and correctness, and recent efforts to quantify the energy efficiency and environmental impact of model inference.

\subsection{Small Language Models for Code Generation}
Small Language Models (SLMs)—typically ranging from hundreds of millions to tens of billions of parameters—have emerged as practical and efficient alternatives to Large Language Models (LLMs) for code generation, particularly in resource-constrained and local deployment scenarios \cite{hasan2025assessing, wang2025comprehensive}. Both general-purpose \textit{(model that was not particularly trained for code generation)} and code-specialized SLMs have proven effective for program synthesis. Code-specialized models such as CodeLlama \cite{roziere2023code}, Qwen3-Coder \cite{yang2025qwen3} benefit from large-scale pretraining on curated code repositories, yielding strong performance on popular benchmarks \cite{chen2021evaluating, austin2021program, hoffmann2022training}. Similarly, a growing body of evidence shows that general-purpose SLMs, trained primarily on natural language with partial exposure to code, can also generate high-quality programs due to their impressive ability for mathematical or logical reasoning \cite{agarwal2025gpt, gemmateam2025gemma3technicalreport}. Unlike general text generation, code generation requires strict syntax, long-range logical dependencies, and executable correctness, making these models highly sensitive to reasoning errors and prompt structure \cite{chen2021evaluating,wang2025codecontests+}. 

The way SLMs are trained and evaluated has direct implications for inference efficiency and environmental impact. Modern SLMs are often instruction-tuned or aligned using execution feedback, which improves correctness but also increases sensitivity to prompt phrasing and intermediate reasoning steps \cite{chen2022program}. At inference time, code generation workloads typically involve long sequences, repeated retries, and multi-turn interactions, especially in real-world developer tools, making inference the dominant contributor to energy use and carbon emissions \cite{ozcan2025quantifying,vartziotis2024carbon}. As a result, small changes in prompt design or reasoning strategy can significantly affect both accuracy and computational cost. This makes SLMs a particularly important setting for studying how prompt engineering and inference-time optimization can impact not only correctness, but also the sustainability of such coding systems.

\subsection{Prompt Engineering} Recent work demonstrates that prompt engineering plays a critical role in improving the correctness, reasoning quality, and efficiency of code generation by large language models \cite{mu2024clarifygpt, cruz2025prompt}. While simpler direct prompting is often ineffective for coding tasks, structured variants that align reasoning with program constructs substantially improve accuracy on coding benchmarks \cite{li2025structured, wei2022chain}. Related work on program-level reasoning shows that explicitly guiding models through structured, executable intermediate representations can further improve correctness and reliability in code generation \cite{chen2022program}. Beyond structured reasoning, careful prompt refinement and iterative prompt design significantly enhance performance even without changing the underlying model \cite{liu2023improving}, whereas multi-turn reasoning and action-based prompting frameworks improve functional correctness at the cost of increased inference overhead \cite{yao2022react}. To mitigate this trade off, automated prompt engineering methods such as evolutionary prompt optimization learn high performing prompts with fewer model calls, achieving competitive or state of the art results across open and closed source code LLMs \cite{taherkhani2024epic}. Adaptive strategies that select prompt techniques based on task complexity further reduce token usage while maintaining accuracy \cite{wang2024selection}. However, recent robustness analyses reveal that code LLM reasoning remains brittle to minor prompt rephrasing, highlighting the need for more stable and reliable prompting methods \cite{roh2025break}. Overall, prompt engineering emerges as a key mechanism for improving both performance and sustainability of code generation systems by reducing unnecessary generation and enabling smaller models to achieve strong results.

\subsection{Energy Efficiency and Environmental Impact of Code Generation Models.}
Recent work increasingly identifies inference—not training—as the dominant contributor to the environmental footprint of large language models \cite{columbia2023ai, patterson2022carbon}. Inference energy consumption scales with model size, output sequence length, and request frequency, making code generation workloads especially costly due to repeated inference and long output sequences \cite{ozcan2025quantifying}. Systematic measurements show that inference energy grows roughly linearly with parameter count, causing large code models to incur disproportionately higher energy and carbon costs compared to smaller alternatives that often suffer only modest accuracy degradation \cite{husom2025sustainable,ashraf2025energy}.

Hardware-level evaluations further reveal that code generation tasks on benchmarks such as HumanEval exhibit relatively low and stable energy per token, as longer outputs amortize inference overhead; however, this characteristic also limits the relative efficiency gains achievable through quantization compared to short-answer reasoning tasks \cite{husom2025sustainable}. Studies of production systems, including GitHub Copilot, demonstrate that the majority of per-task emissions stem from model inference rather than from executing the generated code itself \cite{vartziotis2024carbon}. Moreover, parallel analyses indicate that generated code does not consistently yield downstream energy savings unless models are explicitly guided toward energy-efficient implementations, constraining the ability of code quality improvements to offset inference costs \cite{vartziotis2024learn}.

To mitigate the growing inference-related footprint of code generation models, a range of optimization strategies have been proposed. These include low-bit quantization \cite{giagnorio2025quantizing}, knowledge distillation \cite{su2024distilled}, caching and efficient serving mechanisms \cite{zhu2025towards}, and prompt optimization techniques that reduce unnecessary token generation without compromising correctness \cite{rubei2025prompt}. Complementary approaches such as carbon-aware scheduling and inference frameworks dynamically adapt generation behavior to grid carbon intensity, achieving significant emission reductions without affecting model accuracy or code quality \cite{ozcan2025quantifying,li2024sprout}. Accurate carbon estimation models for LLM inference further enable principled evaluation and optimization of deployment strategies \cite{fu2025llmco2}. Despite these advances and the growing popularity of SLMs, prior work does not systematically analyze how SLMs perform under sustainability metrics such as energy consumption and carbon emissions, and how they differ across different prompting settings during code generation. As a result, the sustainability trade-offs introduced by model scale and reasoning behavior remain largely unexplored.

%% file: experimental-setup_new.tex
\section{Research Methodology}
This section outlines the methodological design of our study for evaluating the trade-offs between code generation performance, computational efficiency, and environmental sustainability.

\subsection{Research Question (RQ)}

The overarching objective of this work is to empirically investigate how different prompting strategies and model characteristics influence both code generation performance and environmental sustainability. Specifically, this study seeks to understand the trade-offs between accuracy, efficiency, and carbon impact when employing SLMs for code generation. We formulate the following research questions (RQs):

\noindent\textbf{\textit{RQ1. How do code generation accuracy and carbon emissions scale across Small Language Model (SLM) parameter sizes when evaluated on standardized datasets?}}

This question examines the trade-off between model performance (measured by Pass@1) and environmental cost (measured in kg of CO$_2$). It aims to identify whether higher-performing models necessarily incur greater carbon emissions, or if more efficient alternatives exist.

\noindent\textit{\textbf{RQ2: What is the impact of various prompting strategies on the trade-off between generation performance, CO2 emissions, and power consumption across different model scales?}}

Here, we analyze how various prompting approaches affect model performance and resource usage. This question seeks to determine whether certain strategies provide a better balance between accuracy and sustainability.

\noindent\textit{\textbf{RQ3: To what extent does the underlying hardware architecture and grid carbon intensity (GCI) influence the energy and emission profiles of automated code generation during inference?}}

This question explores whether and how the underlying hardware \textit{(where the inference is done)} influences the overall power consumption and $CO_2$ emission on code generation tasks.

\noindent\textit{\textbf{RQ4: What are the correlations between token usage, inference time, and computational costs (energy/emissions) when generalized across diverse datasets and evaluation metrics?}}

This research question examines how token usage, inference time, and computational costs (in terms of energy consumption and emissions) relate to one another across a variety of datasets and evaluation metrics, exploring whether trade-offs or consistent patterns emerge.
\subsection{Dataset}
We conduct our evaluation using two widely adopted code generation benchmarks: HumanEval+ and MBPP+. HumanEval+ is an enhanced version of the original HumanEval benchmark \cite{chen2021evaluating} and consists of 164 programming problems, each provided with a function signature, natural language description, unit tests, and reference implementation. We use the full HumanEval+ dataset in our experiments. In addition, we evaluate on the MBPP+ benchmark, which extends the original MBPP dataset \cite{austin2021program} with improved test coverage and clearer problem specifications. From MBPP+, we select a subset of 150 problems. To ensure reproducibility and avoid ordering bias, the problems are randomly shuffled using a fixed seed, and the first 150 samples are selected. 


\subsection{Small Language Models}
Table~\ref{tab:models} lists the 11 open-source language models evaluated in this study. Since no established literature provides strict thresholds for categorizing small language models, for brevity, we group them into three categories—\emph{tiny} (1B--3B), \emph{small} (7B--16B), and \emph{medium} (20B--34B)—based on parameter count. This heuristic enables a systematic comparison across models of different scales and corresponds to practical deployment scenarios, ranging from low-resource to high-capacity inference \cite{hoffmann2022training}. We refer to a model in this study by \textbf{MODEL\_NAME:\#PARAM} (e.g., GPT-OSS:20B refers to the 20B variant of GPT-OSS).

\input{tables/models}

The selected models were chosen to maximize coverage along three orthogonal axes that are central to our study: \emph{model size}, \emph{training objective}, and \emph{reasoning capability}. First, spanning three size regimes allows us to examine how carbon efficiency and accuracy scale with model capacity while avoiding extremely large models whose inference costs would dominate sustainability measurements and obscure comparative trends. Second, we intentionally include both general-purpose (e.g., \texttt{GPT-OSS:20B}) and code-specialized (e.g., \texttt{Qwen3-Coder:30B}) models at each scale to disentangle the effects of domain specialization from raw parameter count. This design enables direct comparisons, such as general-purpose versus coder models at similar sizes. Third, the model set explicitly includes both reasoning-capable and non-reasoning variants. Recent models increasingly incorporate architectural or training modifications aimed at improved reasoning, often at the cost of higher token usage. Including both classes allows us to evaluate whether these reasoning capabilities translate into better accuracy–emission trade-offs under realistic inference settings.

\subsection{Prompting Strategies}
\label{sec:prompting-strategies}

We evaluate six prompting strategies representing major approaches in the literature:

\begin{enumerate}
    \item \textit{Direct.} Direct prompting provides a straightforward problem description and requests code generation. It serves as the baseline approach, with minimal computational and reasoning overhead.

    \item \textit{Chain-of-Thought (CoT) \cite{wei2022chain}.} Chain of Thought augments the prompt with instructions to decompose the problem into intermediate reasoning steps before producing the final code, encouraging structured problem-solving.
    
    \item \textit{Program-of-Thought (PoT) \cite{chen2022program}.} Program of Thought similarly decomposes the problem but additionally executes generated code during inference to validate intermediate results. In our setup, PoT performs syntactic verification of generated code before outputting the final solution.
    
    \item \textit{Self Consistency (CoT-sampled) \cite{wang2022self}.} Self Consistency generates multiple independent CoT solutions (here, 5 samples) and selects the most consistent answer through voting or consensus. If all samples differ, the first complete solution is chosen.
    
    \item \textit{Least-to-Most \cite{zhou2022least}.} This approach adopts a two-stage approach: first decomposing the problem into sub-problems, then solving these sequentially while propagating intermediate results.
    
    \item \textit{ReAct \cite{yao2022react}.} ReAct implements an iterative loop of reasoning, action, and observation, allowing the model to plan, execute, and refine solutions across multiple steps. Since MBPP+ and HumanEval+ consist primarily of basic programming problems, we employ ReAct with a syntax validation tool that checks whether generated code is syntactically correct.
\end{enumerate}

\subsection{Evaluation Framework Setup}
Experiments are conducted using a standardized evaluation framework that integrates prompting strategies, model inference, and sustainability tracking.

\noindent\textit{\textbf{Prompting Framework.}} All prompting strategies used in this study are implemented using the DSPy framework \cite{khattab2023dspy}, which provides a consistent abstraction layer across diverse prompting paradigms. DSPy structures each task according to a signature-based schema, where input fields (e.g., problem descriptions and constraints) and output fields (e.g., generated code and reasoning steps) are explicitly labeled and serialized into a canonical prompt. For reasoning strategies, including CoT and ReAct, DSPy automatically introduces dedicated reasoning fields, manages intermediate steps, and enforces structured output formatting.

Additionally, we adopt DSPy to mitigate the variability inherent in ad-hoc prompting, which can produce inconsistent outputs, malformed code, or verbose responses that do not reflect realistic developer workflows \cite{lemos2025time}. By enforcing canonical, machine-parsable code outputs, DSPy ensures that only syntactically correct and well-structured code is accepted, mirroring real-life scenarios where generated code must compile or be integrated without extensive manual correction. While parsing errors can still occur, this structured approach substantially reduces inconsistencies across models and backends, enabling a more reliable and reproducible assessment of both accuracy and environmental metrics

\begin{figure}
    \centering
    \includegraphics[width=\linewidth]{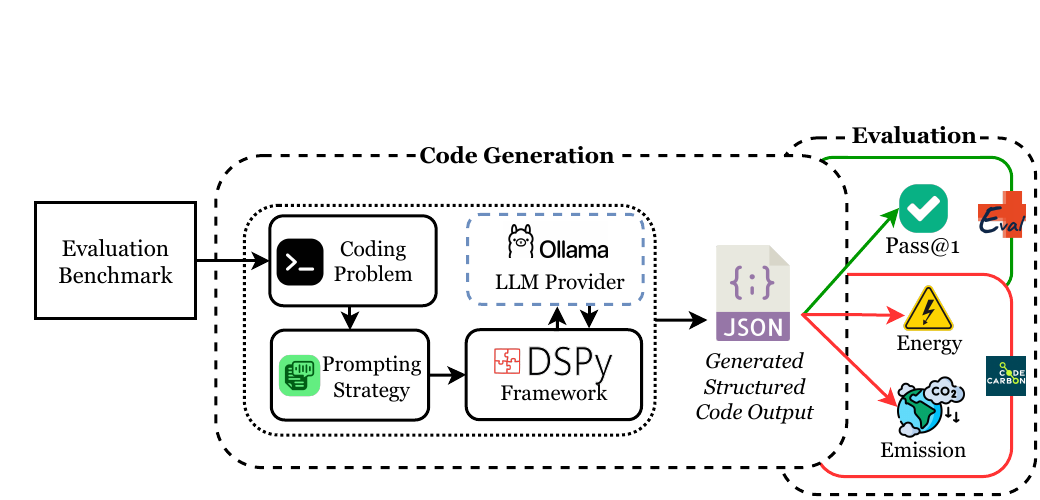}
    \caption{Evaluation framework to benchmark LLMs.}
    \label{fig:sustainability-framework}
\end{figure}

\noindent\textit{\textbf{LLM Provider.}} Model inference is served through Ollama\footnote{\url{https://github.com/ollama/ollama}}, which provides unified access to open-source models. Figure~\ref{fig:sustainability-framework} illustrates the overall evaluation workflow.

\noindent\textit{\textbf{Hardware Configuration.}} To capture both hardware- and region-specific effects, experiments are conducted on two distinct computing environments. Machine 1 equips a single Xeon Silver CPU and an NVIDIA A100 GPU. Machine 2 is a high-performance system with 96 Xeon Gold CPUs and an NVIDIA L40S GPU and is used to study the impact of hardware scaling and regional differences in grid carbon intensity. This dual-environment design allows the evaluation of the sensitivity of energy consumption and carbon emissions to both computational scale and electricity source, complementing analyses of accuracy and token efficiency.

\subsection{Carbon Emission Estimation (CO$_2$eq)}

Equivalent carbon dioxide emission (CO$_2$eq) quantifies the environmental impact of model inference by combining measured energy consumption with the carbon intensity of the electricity used. In this work, CO$_2$eq is computed using the CodeCarbon framework\footnote{\url{http://codecarbon.io}}, which estimates emissions as the product of total energy consumed and the carbon intensity of the local electricity grid. Formally, for a given inference run, CO$_2$eq is defined as:
\begin{equation}
\text{CO$_2$eq} = E \times \gamma,
\end{equation}
where $E$ denotes the total energy consumed (by the compute machine) during inference (in kWh), and $\gamma$ represents the carbon intensity of electricity (in kgCO$_2$/kWh).

\noindent\textit{{\textbf{Energy Consumption ($E$).}}}
CodeCarbon measures energy consumption by aggregating power usage across all active hardware components, including CPU, GPU, and RAM, over the duration of execution. Instantaneous power measurements are integrated over time to obtain the total energy consumed per problem or per experiment.

\noindent\textit{{\textbf{Carbon Intensity ($\gamma$).}}}
The carbon intensity $\gamma$ reflects the amount of CO$_2$ emitted per unit of electricity generated and is derived from region-specific electricity grid data. CodeCarbon uses country and province-level or cloud-provider-specific carbon intensity values, which are computed as weighted averages of emissions from fossil-fuel-based sources (e.g., coal and natural gas) and low-carbon or renewable sources (e.g., nuclear, solar), based on the underlying energy mix of the local grid.

\subsection{Evaluation Metrics}
To comprehensively evaluate each (model, strategy) configuration, we consider a combination of performance, efficiency, and environmental impact metrics. Let $N$ denote the total number of problems, and let $T_i$, $E_i$, and $P_i(t)$ represent the inference time, energy consumed, and instantaneous power for the $i$-th problem, respectively.

\noindent\textit{\textbf{Pass@1 (\%).}} Pass@1 measures the proportion of problems solved correctly on the first attempt:
\begin{equation}
\text{Pass@1} = \frac{1}{N} \sum_{i=1}^{N} \mathbb{1}(\text{correct}_i) \times 100,
\end{equation}
where $\mathbb{1}(\cdot)$ is an indicator function that equals 1 if the model's generated solution for problem $i$ is correct and 0 otherwise.

\noindent\textit{\textbf{Average CO$_2$ Emission (kg).}  }
The average carbon emission or equivalent carbon emission (CO$_2$eq) per problem accounts for the device energy consumption $E_i$ and the local electricity grid carbon intensity $\gamma_i$:
\begin{equation}
\text{Avg. CO$_2$/CO$_2$eq} = \frac{1}{N} \sum_{i=1}^{N} \left( E_i \times \gamma_i \right),
\end{equation}
where $\gamma_i$ is in kgCO$_2$/kWh.

\noindent\textit{\textbf{Average Inference Time (s).}}
The average inference time per problem is given by:
\begin{equation}
\text{Avg. Inference Time} = \frac{1}{N} \sum_{i=1}^{N} T_i.
\end{equation}

\noindent\textit{\textbf{Average Total Tokens.}}
The average total token count (prompt + completion) per problem is computed as:
\begin{equation}
\text{Avg. Tokens} = \frac{1}{N} \sum_{i=1}^{N} (t_i^{\text{prompt}} + t_i^{\text{completion}}).
\end{equation}

\noindent\textit{\textbf{Average Energy Consumption (kWh).}  }
The average energy consumed per problem during inference is computed as:
\begin{equation}
\text{Avg. Energy} = \frac{1}{N} \sum_{i=1}^{N} E_i,
\end{equation}

where $E_i$ denotes the total energy consumed for problem $i$, measured in kWh and obtained directly from CodeCarbon’s \textit{energy\_consumed} metric (the sum of CPU, GPU, and RAM energy).

%% file: tables/models.tex
\begin{table}[ht]
\centering
\caption{Language models evaluated in this study and their primary characteristics.}
\label{tab:models}
\resizebox{\columnwidth}{!}{%
\begin{tabular}{lcccc}
\toprule
\textbf{Model} & \textbf{Category} & \textbf{Objective} & \textbf{Reasoning} & \textbf{Parameters} \\
\midrule

Llama3.2:1b \cite{grattafiori2024llama3herdmodels}           & Tiny   & General-purpose & No & 1B \\
Qwen3:1.7b \cite{yang2025qwen3}             & Tiny   & General-purpose & Yes & 1.7B \\
StarCoder2:3b  \cite{lozhkov2024starcoder}        & Tiny  & Code-specialized   & No  & 3B \\

\midrule

CodeGemma:7b \cite{codegemmateam2024codegemmaopencodemodels}          & Small  & Code-specialized   & No  & 7B \\
Qwen2.5-Coder:7b \cite{hui2024qwen2}       & Small  & Code-specialized   & No  & 7B \\
Phi4:14b \cite{abdin2024phi4technicalreport}              & Small & General-purpose & Yes & 14B \\
DeepSeek-Coder-V2:16b \cite{zhu2024deepseekcode}  & Small & Code-specialized   & Yes  & 16B \\

\midrule
GPT-OSS:20b \cite{agarwal2025gpt}          & Medium & General-purpose & Yes & 20B \\
Gemma3:27b \cite{gemmateam2025gemma3technicalreport}            & Medium & General-purpose & No & 27B \\
Qwen3-Coder:30b \cite{yang2025qwen3}        & Medium & Code-specialized   & Yes & 30B \\
CodeLlama:34b \cite{roziere2023code}         & Medium & Code-specialized   & No  & 34B \\

\bottomrule
\end{tabular}%
}
\end{table}

%% file: results_new.tex
\section{Results}

We report the results of our study in this section, organized by research question. We examine the relationships between accuracy and emissions across models, the trade-offs introduced by different prompting strategies, the influence of hardware and regional carbon intensity, and the connections between accuracy and sustainability metrics.

\subsection{RQ1: Accuracy-Emissions Relationship Across SLMs}

\noindent\textbf{Approach.} We prompt all models directly with the problem statements to generate code solutions. Generated code was evaluated against the provided test cases to compute Pass@1 rates in both HumanEval+ and MBPP+ datasets, while CO$_2$ emissions were logged for each generation.

\begin{figure}
    \centering
    \includegraphics[width=1\linewidth]{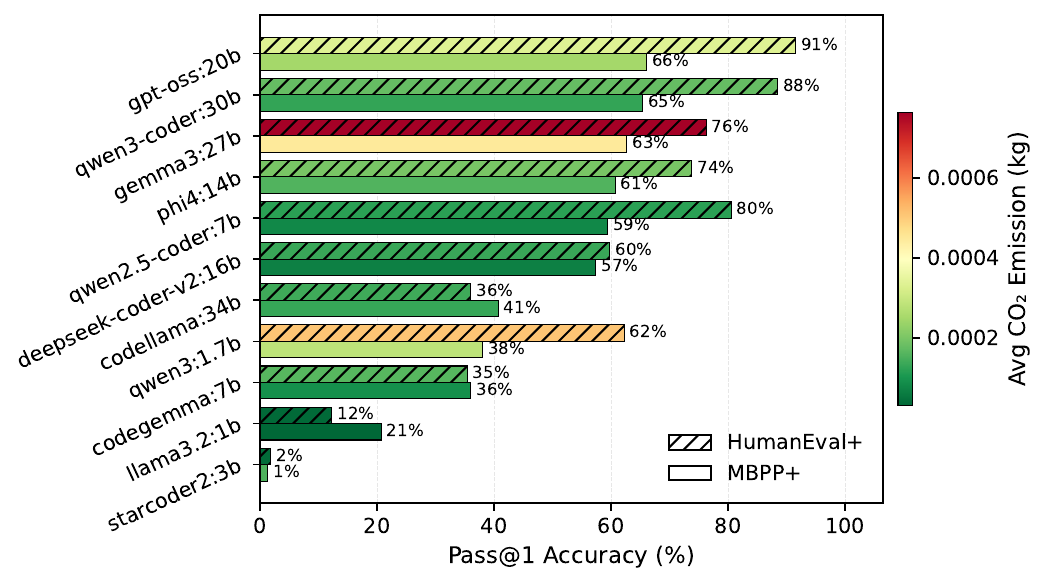}
    \caption{Comparison of Pass@1 accuracy and CO$_2$ emissions on MBPP+ and HumanEval+ across models.}
    \label{fig:model-heatmap-acc}
\end{figure}

\noindent\textbf{Results.} Figure~\ref{fig:model-heatmap-acc} illustrates the trade-offs between model accuracy and emissions.

\noindent\textit{\textbf{Code Generation Accuracy.}} In both datasets, code generation accuracy does not scale monotonically with parameter count. While GPT-OSS:20B achieves the highest average Pass@1 91\% on HumanEval+ and 66\% on MBPP+, it outperforms its closest larger competitor Qwen3-Coder:30B by 3.4\% and 1.5\% on HumanEval+ and MBPP+, respectively. Similarly, Qwen2.5-Coder:7B outperforms larger models like Gemma3:27B and Phi4:14B models. A model as small as Qwen3:1.7B outperforms 20$\times$ larger model like CodeLlama 34B due to its capability of reasoning. However, similar tiny models such as StarCoder2:3b and Llama3.2:1B perform poorly on both HumanEval+ and MBPP+. This degradation is not solely attributable to limited code generation capability. In practice, these models exhibit compatibility issues with the DSPy framework, frequently failing to produce structured outputs that conform to the expected DSPy signatures. An illustrative example in Listing \ref{lst:dspy-json-failure} shows that StarCoder2:3B outputs a few "zeros" for \texttt{HumanEval/24} rather than a valid Python function. We discuss this in-depth in Section \ref{sec:parsing-errors}. 

\begin{lstlisting}[language={}, frame=single, label={lst:dspy-json-failure}, caption={Example of invalid output by StarCoder2.}]
{
        "problem_id": "HumanEval/24",
        "code": "000000",
        "prompt_strategy": "direct",
}
\end{lstlisting}

In addition, we observe that Qwen3:1.7B exhibits substantially higher average token usage (2,690 tokens) than all other evaluated models. Applying the Tukey Interquartile Range (IQR) criterion ($Q3 + 1.5 \times IQR \approx 1777$ tokens), Qwen3:1.7B is identified as a statistical outlier on both datasets and excluded from subsequent analyses to avoid skewing aggregate results. Accordingly, for the remaining research questions, we exclude Llama3.2:1B and StarCoder2:3B due to unreliable output formatting, and Qwen3:1.7B due to disproportionate token usage.

\noindent\textit{\textbf{CO$_2$ Emission Profile.}} Average Per-query CO$_2$ emissions fall within a narrow range (approximately $3.11 \times 10^{-5}$--$7.63 \times 10^{-4}$ kg) and show weak dependence on model size. For example, Gemma3:27B exhibits the highest emissions ($6.04 \times 10^{-4}$kg), whereas both larger models like CodeLlama:34B and the more accurate Qwen3-Coder:30B averages around $1.5 \times 10^{-4}$ kg on both datasets, suggesting that inference efficiency of a model and runtime configuration dominate emission profile. Similarly, on the HumanEval+ dataset, the Qwen3-1.7B model emitted substantially higher CO$_2$ due to its significantly greater token consumption during the thinking process, exceeding the emission profiles of all other models except Gemma3:27B.

\noindent\textit{\textbf{Model Specialization.}} We did not find any conclusive improvement in either accuracy or emission of code-specialized models over traditional models. For example, Qwen3:1.7B--a tiny general-purpose SLM outperformed CodeGemma:7B by 77\% and 5\% on HumanEval+ and MBPP+ datasets respectively while its emission profile was also high. On the other side, Qwen2.5-coder:7b--a small code specialized model outperformed both Gemma3:27B (general-purpose model) by 5.2\% on HumanEval+ dataset.

\begin{summarybox}
\textbf{RQ1 Summary. }
Code generation accuracy does not scale monotonically with model size, while per-query CO$_2$ emissions remain weakly correlated with parameter count. Well-trained general-purpose models can also be as effective as code-specialized models. However, tiny models often struggle with structured outputs.
\end{summarybox}

\subsection{RQ2: Impact of Prompting Strategies on Accuracy, Emissions, Energy, and Token Usage}
\noindent\textbf{Design.} We evaluate six prompting strategies defined in Section \ref{sec:prompting-strategies}. For each strategy, the generated code is assessed against test cases to calculate Pass@1 accuracy, CO$_2$ emissions, average power consumption, and token usage are logged to evaluate computational and environmental cost. For a fair comparison, all strategy-model combination was evaluated under the same hardware configuration (Machine 1).

\noindent\textbf{Results.} Table~\ref{tab:strategy_performance} shows trade-offs between accuracy and environmental cost across prompting strategies. Self-Consistency achieves the highest mean Pass@1 accuracy (65.75\%), outperforming all other strategies on both HumanEval+ and MBPP+. However, this gain comes at a substantial computational cost, with the highest average energy consumption of 0.001806~kWh and CO$_2$ emissions of 0.001133~kg (4.9$\times$ compared to CoT), driven by its requirement to sample and evaluate multiple reasoning trajectories.

CoT offers a notably more efficient accuracy--cost balance. While its mean accuracy (63.94\%) is only about 2.7\% lower than Self-Consistency, it reduces energy usage by nearly 80\% and CO$_2$ emissions by a similar margin. This demonstrates that explicit intermediate reasoning can yield competitive performance without the substantial overhead of multi-sample aggregation. Direct prompting incurs the lowest environmental cost among all strategies (0.000338~kWh, 0.000212~kg) and achieves an average Pass@1 of 61.84\%. However, we note that Direct prompting in the DSPy framework often consumes more prompt tokens than CoT. We discuss this further in Section \ref{sec:direct-vs-cot}. 

More complex reasoning frameworks—LtoM, PoT, and ReAct do not consistently translate increased computational effort into proportional accuracy gains. LtoM and ReAct incur some of the highest token counts (5,006 and 5,889 tokens on average, respectively) and energy consumption (up to 0.001677~kWh for LtoM), yet their mean accuracies remain comparable or worse than Direct prompting. These results indicate diminishing returns from more elaborate decomposition or tool-based reasoning, particularly when applied to small-scale language models for straightforward code generation.

\input{tables/strategy-performance}

\begin{figure}[ht]
    \centering
    \includegraphics[width=1\linewidth]{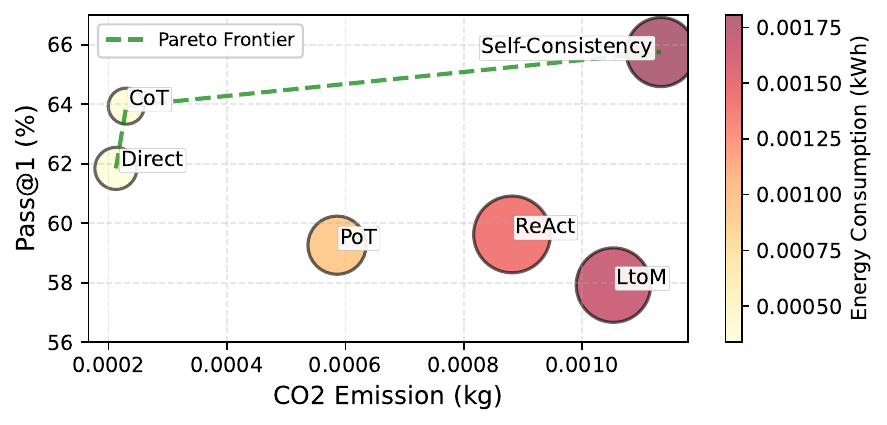}
    \caption{Bubble chart of mean Pass@1 accuracy versus energy consumption for different prompting strategies. Bubble size represents average token count and color represents average CO$_2$ emissions.}
    \label{fig:strategy-comparison}
\end{figure}

\noindent\textbf{\textit{Trade-offs.}} Figure~\ref{fig:strategy-comparison} visualizes the trade-off between Pass@1 accuracy and CO$_2$ emissions across prompting strategies. Chain-of-Thought (CoT) lies closest to the Pareto frontier, achieving strong accuracy with relatively low emissions, while Direct prompting minimizes environmental cost at the expense of accuracy. Self-Consistency delivers the highest Pass@1 but incurs substantially higher energy consumption and CO$_2$ emissions due to multi-sample aggregation, placing it in a high-accuracy, high-cost regime.

More complex strategies, such as ReAct, Least-to-Most, and Program-of-Thought, are dominated in this space, as they incur higher token usage and emissions without corresponding accuracy gains. Bubble size and color further indicate that increased token generation closely influences energy and CO$_2$ cost (e.g., bigger bubbles are more aligned to the right side of the plot while the color gets more intensified). Overall, the results suggest that for small language models, simpler prompting strategies, particularly CoT, provide the most favorable accuracy–efficiency trade-off for code generation tasks.

\begin{summarybox}
\textbf{RQ2 Summary.}  
Prompting strategies introduce clear trade-offs between code generation accuracy and environmental cost. Strategies like Self-Consistency achieve the highest accuracy but incur substantial energy and emissions, while simpler methods like CoT maintain competitive performance with much lower environmental impact.
\end{summarybox}

\subsection{RQ3: Influence of Underlying Hardware and Region}

\noindent\textbf{Approach.} To answer this RQ, we evaluate how hardware configuration and regional electricity characteristics jointly influence the carbon emissions of model inference. We conduct experiments on two research computers deployed in different Canadian provinces: Machine 1 (M\#1) in Alberta and Machine 2 (M\#2) in Ontario. The machines differ in both processor configuration and accelerator class; notably, Machine 2 employs higher-power components and therefore exhibits a higher expected energy consumption profile. The detailed hardware specifications are summarized in Table \ref{tab:hardware}. For each machine, we run identical inference workloads on the MBPP+ dataset using the same models and prompting strategies, ensuring that performance and workload characteristics remain comparable across settings. Since CoT emerged as the prompting strategy closest to the Pareto frontier, we considered this strategy to represent this RQ. Nevertheless, data for all the other strategies are still available in the replication package,

\input{tables/hardware-details}

\begin{figure}[ht]
    \centering
    \includegraphics[width=1\linewidth]{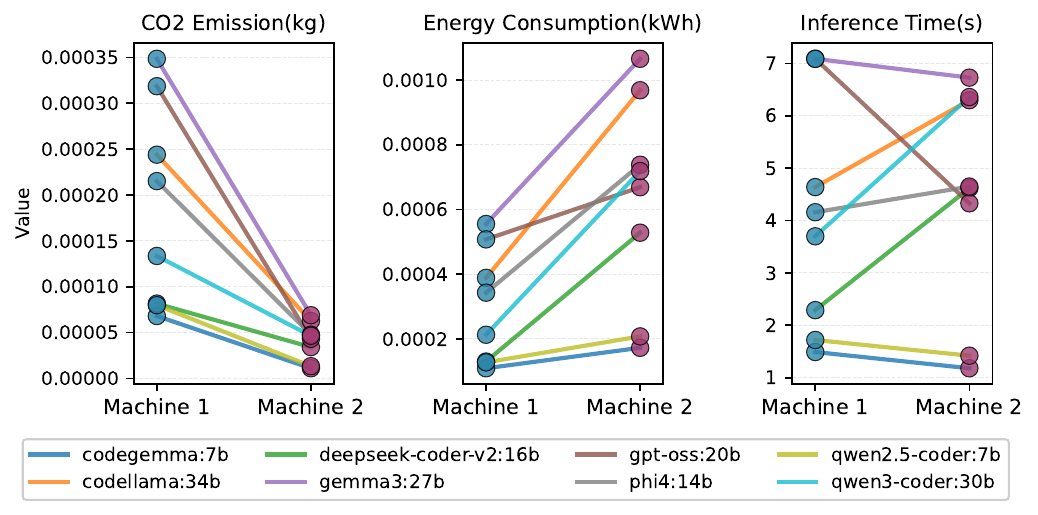}
    \caption{Comparison of CO$_2$ emission, energy consumption, and inference time of different models in Machine 1 and Machine 2 when Chain-of-Thought is used.}
    \label{fig:hardware-comparison}
\end{figure}

\noindent\textbf{Results.} In RQ2, we observe how energy consumption and CO$_2$ emission change proportionally when the experiments were conducted on a single machine (i.e., Machine 1). However, Figure~\ref{fig:hardware-comparison} shows that carbon emissions are not always proportional to energy consumption across different regions. For example, Machine 2 consumes substantially more energy per inference, with an average energy usage that is $113.54\%$ higher than Machine~1, but its carbon emission profile is significantly lower, achieving a $78\%$ reduction in average CO$_2$ output.

\noindent\textbf{\textit{Grid Carbon Intensity (GCI).}} The discrepancy in emission is primarily driven by differences in regional grid carbon intensity. Alberta’s electricity grid is dominated by fossil-fuel-based generation, particularly natural gas and coal, which have high carbon intensities (e.g., 743~kgCO$_2$/MWh for natural gas and 995~kgCO$_2$/MWh for coal\footnote{\url{https://mlco2.github.io/codecarbon/methodology.html}}). In contrast, Ontario’s grid relies on low-carbon energy sources such as nuclear and hydroelectric power, with carbon intensities of 29~kgCO$_2$/MWh and 26~kgCO$_2$/MWh, respectively. Based on the regional energy mixes, the estimated grid carbon intensity calculated by \texttt{codecarbon} library is approximately 0.627~kgCO$_2$/kWh for Alberta and 0.065~kgCO$_2$/kWh for Ontario, corresponding to an $89.6\%$ lower GCI in Ontario.

\input{tables/grid-intensity}
Despite near-identical model performance across the two machines---with average Pass@1 differing by only $0.71\%$---the emissions outcomes differ substantially. As shown in Table~\ref{tab:grid_intensity_comparison}, although Machine 2 exhibits higher electricity and token usage, yet the lower-carbon electricity mix in Ontario dominates the emissions calculation. Consequently, the average CO$_2$ emissions on Machine~2 (0.000041~kg) are markedly lower than those on Machine~1 (0.000186~kg), despite higher absolute energy consumption. For a counterfactual analysis, if Machine 2 had been operated under Alberta’s electricity mix, its estimated CO$_2$ emissions would increase to 0.0004~kg—over twice that of Machine~1. This highlights grid carbon intensity as a first-order determinant of inference-time emissions, often outweighing differences in token usage or energy consumption.

\begin{summarybox}
    \textbf{RQ3 Summary. } The results show that grid carbon intensity is the primary determinant of inference-time CO$_2$ emissions, dominating the impact of hardware energy consumption. Differences in regional electricity carbon intensity lead to large emission disparities even when hardware energy usage is higher.
\end{summarybox} 

\begin{figure*}[ht]
    \centering
    \includegraphics[width=0.95\linewidth]{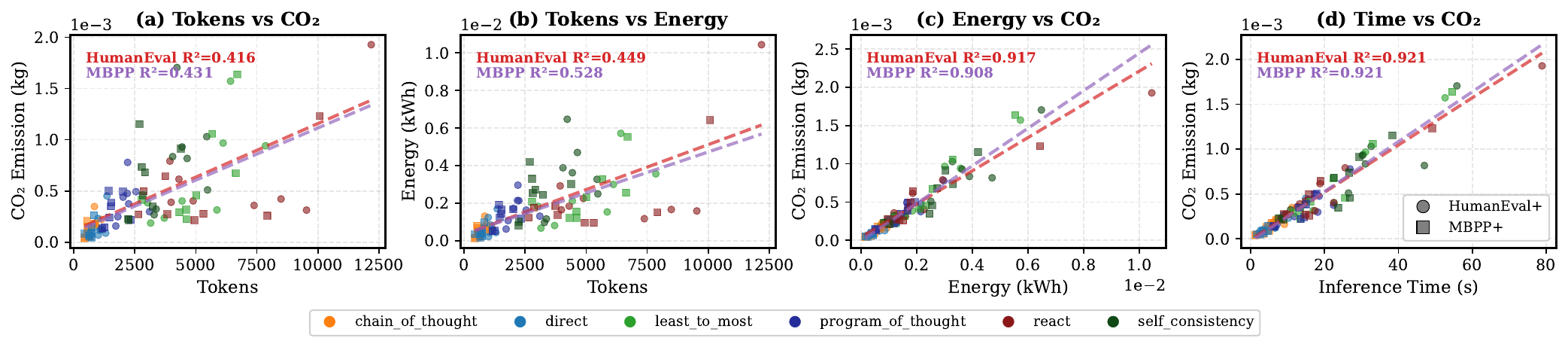}
    \caption{Relationship between different sustainability factors across different prompting strategies.}
    \label{fig:r2-comparison}
\end{figure*}

\subsection{RQ4: Relationship Between Token Usage and Sustainability Factors}

\noindent\textbf{Approach.}
We analyze the relationships between token usage, inference time, energy consumption, and CO$_2$ emissions using the coefficient of determination (R$^2$). We also investigate Pearson correlation coefficients between both performance metrics and sustainability metrics, which are computed per dataset and aggregated across hardware configurations (M\#1, M\#2) to assess whether observed trends generalize across evaluation settings. In addition, we report R$^2$ values for key variable pairs to quantify the extent to which token usage and inference time explain variance in energy use and emissions.

\noindent\textbf{Results.} Figure~\ref{fig:r2-comparison} shows that inference time is the strongest predictor of CO$_2$ emissions across both datasets, exhibiting an almost perfect correlation. Energy consumption displays similarly high explanatory power, accounting for over 91\% of the variance in emissions, confirming that emissions are primarily driven by how long and how intensively hardware is utilized during inference.

In contrast, token usage exhibits only moderate correlations with both energy consumption and CO$_2$ emissions (R$^2$ = 0.41–0.52). This indicates that token counts alone capture only part of the sustainability-related variability. For instance, two inference runs with comparable token counts can result in substantially different energy usage if one completes faster due to more efficient decoding, better hardware utilization, or differences in prompting structure. Conversely, a run generating more tokens may consume less energy if executed more efficiently. These observations highlight that token-based workload metrics may not be sufficient for accurately characterizing environmental impact. Instead, inference time may provide more reliable indicators.

\noindent\textbf{\textit{Correlation between performance and sustainability metrics.}} Table \ref{tab:correlation_matrices_merged} also shows that the code generation metric (Pass@1) shows no meaningful correlation with any sustainability factor, indicating that higher performance does not inherently entail higher energy or emissions. Moreover, energy consumption and CO$_2$ emission are weakly correlated. This is due to the variations in the grid carbon intensity (GCI) (as observed in RQ3) dominating the overall emission profile, further decoupling workload from actual environmental cost.

\begin{summarybox}
\textbf{RQ4 Summary.}  
Environmental impact during model inference is primarily determined by runtime and energy consumption, which are strongly predictive of CO$_2$ emissions across datasets and prompting strategies, while model performance (Pass@1) shows no meaningful correlation with sustainability metrics.
\end{summarybox}
\input{tables/correlation_table}

%% file: tables/strategy-performance.tex
\begin{table}[t]
\centering
\caption{Comparison of prompting strategies in terms of Pass@1 accuracy, energy consumption, and CO$_2$ emissions on HumanEval+ and MBPP+.}

\label{tab:strategy_performance}
\resizebox{\columnwidth}{!}{
\begin{tabular}{llcccc}
\toprule
\textbf{Strategy} & \textbf{Dataset} & \textbf{Pass@1 (\%)} & \textbf{CO$_2$ (kg)} & \textbf{Energy (kWh)} & \textbf{Tokens} \\
\midrule
CoT & HumanEval+ & 69.21 & 0.000274 & 0.000436 & 853.98 \\
                 & MBPP+      & 58.66 & 0.000186 & 0.000297 & 560.78 \\
                 & \textbf{Mean} & \textbf{63.94} & \textbf{0.000230} & \textbf{0.000366} & \textbf{707.38} \\
\midrule
Direct & HumanEval+ & 67.68 & 0.000254 & 0.000405 & 1075.56 \\
                 & MBPP+      & 56.00 & 0.000170 & 0.000271 & 681.18 \\
                 & \textbf{Mean} & \textbf{61.84} & \textbf{0.000212} & \textbf{0.000338} & \textbf{878.37} \\
\midrule
LtoM & HumanEval+ & 61.51 & 0.001048 & 0.001669 & 5035.90 \\
                 & MBPP+      & 54.33 & 0.001058 & 0.001685 & 4976.43 \\
                 & \textbf{Mean} & \textbf{57.92} & \textbf{0.001053} & \textbf{0.001677} & \textbf{5006.16} \\
\midrule
PoT & HumanEval+ & 64.03 & 0.000666 & 0.001060 & 2170.76 \\
                 & MBPP+      & 54.50 & 0.000506 & 0.000807 & 1545.85 \\
                 & \textbf{Mean} & \textbf{59.26} & \textbf{0.000586} & \textbf{0.000933} & \textbf{1858.31} \\
\midrule
ReAct & HumanEval+ & 64.33 & 0.001031 & 0.001642 & 6610.77 \\
                 & MBPP+      & 54.92 & 0.000732 & 0.001167 & 5167.92 \\
                 & \textbf{Mean} & \textbf{59.62} & \textbf{0.000882} & \textbf{0.001404} & \textbf{5889.34} \\
\midrule
Self-Consistency & HumanEval+ & 71.34 & 0.001309 & 0.002086 & 4244.56 \\
                 & MBPP+      & 60.17 & 0.000957 & 0.001526 & 2816.19 \\
                 & \textbf{Mean} & \textbf{65.75} & \textbf{0.001133} & \textbf{0.001806} & \textbf{3530.37} \\
\bottomrule
\end{tabular}
}
\end{table}

%% file: tables/hardware-details.tex



\begin{table}[t]
\centering
\caption{Hardware configurations used in the experiments.}
\label{tab:hardware}
\resizebox{\columnwidth}{!}{%
\begin{tabular}{lcccc}
\toprule
\textbf{Identifier} & \textbf{CPU (TDP)} & \textbf{GPU (TDP)} & \textbf{RAM} & \textbf{Region} \\
\midrule
Machine 1 & Xeon Silver 4316 (150W) & A100 40GB (250W) & 64 GB & Alberta \\
Machine 2 & Xeon Gold 6442Y (225W) & L40S 48GB (350W) & 250 GB & Ontario \\
\bottomrule
\end{tabular}
}
\end{table}

%% file: tables/grid-intensity.tex
\begin{table}[ht]
\centering
\caption{Comparison of different evaluation metrics on MBPP+ across two machines in different regions.}
\resizebox{\columnwidth}{!}{
\begin{tabular}{lccc}
\toprule
\textbf{Metric} & \textbf{M\#1 (Alberta)} & \textbf{M\#2 (Ontario)} & \textbf{Diff.(\%)} \\
\midrule
Avg. Pass@1 (\%) & 58.66 & 58.25 & -0.71 \\
Avg. Total Tokens & 561 & 601 & +7.23 \\
Avg. Energy (kWh) & 0.000297 & 0.000634 & +113.54 \\
Estimated GCI (CO$_2$eq/kWh) & $\approx 0.627$ & $\approx 0.065$ & -89.69 \\
Avg. CO$_2$ Emissions (kg) & 0.000186 & 0.000041 & -77.97 \\
\bottomrule
\end{tabular}
}
\label{tab:grid_intensity_comparison}
\end{table}

%% file: tables/correlation_table.tex
\begin{table}[htbp]
\centering
\caption{Correlation matrices for performance and sustainability metrics across datasets (merged across hardware configurations). Values show Pearson correlation coefficients.}
\label{tab:correlation_matrices_merged}

\resizebox{\columnwidth}{!}{
\begin{tabular}{l|ccccc|ccccc}
\toprule
& \multicolumn{5}{c|}{\textbf{HumanEval+}} & \multicolumn{5}{c}{\textbf{MBPP+}} \\
\cmidrule(lr){2-6} \cmidrule(lr){7-11}
 & Pass@1 & CO$_2$ & Tokens & Time & Energy
 & Pass@1 & CO$_2$ & Tokens & Time & Energy \\
\midrule
Pass@1 & -- & 0.041 & -0.164 & 0.012 & -0.067 & -- & 0.038 & -0.183 & -0.022 & -0.113 \\

CO$_2$ & 0.041 & -- & 0.441 & \cellcolor{corrLow}0.652 & 0.308 & 0.038 & -- & 0.458 & \cellcolor{corrMid}0.720 & 0.331 \\

Tokens & -0.164 & 0.441 & -- & \cellcolor{corrMid}0.662 & 0.593 & -0.183 & 0.458 & -- & \cellcolor{corrMid}0.677 & \cellcolor{corrLow}0.626 \\

Time & 0.012 & \cellcolor{corrLow}0.652 & \cellcolor{corrMid}0.662 & -- & \cellcolor{corrTop}0.905 & -0.022 & \cellcolor{corrMid}0.720 & \cellcolor{corrMid}0.677 & -- & \cellcolor{corrTop}0.872 \\

Energy & -0.067 & 0.308 & 0.593 & \cellcolor{corrTop}0.905 & -- & -0.113 & 0.331 & \cellcolor{corrLow}0.626 & \cellcolor{corrTop}0.872 & -- \\
\bottomrule
\end{tabular}
}
\end{table}

%% file: discussion.tex
\section{Discussion}
This section interprets some empirical findings from our results, providing deeper insight into the mechanisms underlying the observed accuracy–sustainability trade-offs. We examine how prompt structure and framework design influence token usage, introduce a normalized accuracy–emissions metric, analyze output-format reliability of SLMs, and discuss broader implications for sustainable code generation systems.

\subsection{Effects of DSPy Prompt Structure on Token Consumption}
\label{sec:direct-vs-cot}
In RQ2 (Table \ref{tab:strategy_performance}), we observed that direct prompting consumes more average tokens than CoT, despite the expectation that CoT should incur higher token usage due to explicit intermediate reasoning. To explain this counterintuitive result, we analyze token usage across four representative models of various sizes, including both reasoning-oriented (DeepSeek-Coder-V2:16B, Qwen3:1.7B) and non-reasoning models (Llama3.2:1B, CodeLlama:34B). 

\begin{figure}[ht]
    \centering
    \includegraphics[width=1\linewidth]{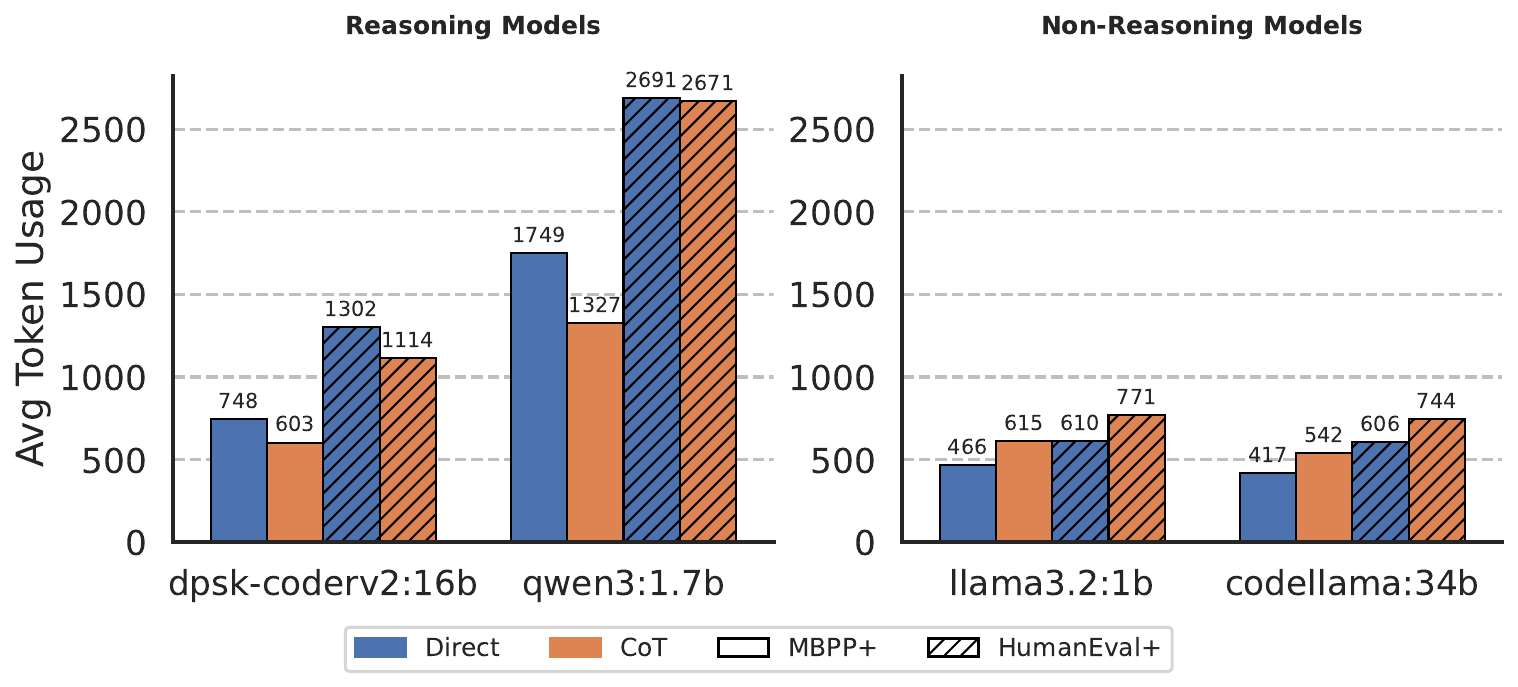}
    \caption{Comparison of average token usage of different reasoning and non-reasoning models.}
    \label{fig:model-token-comparison}
\end{figure}

\noindent\textbf{\textit{Observation.}} As shown in Figure \ref{fig:model-token-comparison}, models that can reason over their generation tend to use more tokens with direct prompting, whereas non-reasoning models consume fewer tokens with direct prompting than with CoT as expected.

This behavior results from the interaction between language model training characteristics and how prompts are constructed in the DSPy framework. Under direct prompting, DSPy’s prediction API\footnote{\url{https://dspy.ai/api/modules/Predict/}} enforces a structured input–output format but does not explicitly constrain generation length within the output field. Reasoning-oriented models, which are often trained to internally deliberate, may implicitly include reasoning within the final answer when no separate reasoning field is provided, leading to longer generations and higher token usage.

In contrast, DSPy's ChainOfThought API\footnote{\url{https://dspy.ai/api/modules/ChainOfThought/}} explicitly introduces a dedicated reasoning field and a fixed instruction to reason step-by-step, followed by a clearly separated final answer field. For reasoning-capable models, this structure constrains where reasoning is expressed, resulting in more concise final answers and, in some cases, lower overall token usage than direct prompting. For non-reasoning models, however, CoT forces the generation of an additional reasoning section that would not otherwise be produced, increasing token usage relative to direct prompting.

Overall, these results indicate that token consumption depends not only on whether reasoning is explicitly requested, but also on how prompting strategies align with a model’s training and inference behaviors. In DSPy, direct prompting does not suppress latent reasoning in reasoning-oriented models, whereas CoT imposes structural constraints that can lead to more efficient token usage.

\subsection{Accuracy-Emission Ratio}

To evaluate the sustainability of different prompting strategies, we introduce the Carbon per Correct Answer (CpCA) metric, which measures the estimated CO$_2$ emissions required to produce one correct solution over both datasets. This metric is directly inspired by the Software Carbon Intensity (SCI) specification \cite{GreenSoftwareFoundation}, which quantifies emissions per functional unit of software work. In our context, the functional unit corresponds to a correctly solved benchmark problem, while emissions are estimated from operational energy use during inference.

\noindent\textbf{\textit{Observation.}} Table~\ref{tab:cpca_machine_comparison} presents CpCA values across multiple prompting strategies and two hardware configurations (defined in Table \ref{tab:hardware}). Lower CpCA values indicate higher sustainability, reflecting a strategy that achieves correctness more efficiently in terms of carbon output. Across both machines, CpCA reveals a consistent pattern: strategies that introduce additional reasoning steps tend to increase emissions faster than they improve accuracy. While methods such as Self-Consistency improve Pass@1 by only a few percentage points, their CpCA increases by over 5 times in both machines.

CpCA also highlights an interaction between prompting strategy and hardware/regional impacts. While Machine 2 exhibits uniformly lower CpCA values due to lower GCI in Ontario compared to Machine 1 in Alberta, the relative ranking of strategies remains largely stable, highlighting that software-level choices—such as prompting design—play a critical role alongside infrastructure in reducing emissions.
\input{tables/carbon-accurate}
\subsection{Output Format Correctness of SLMs}
\label{sec:parsing-errors}

Beyond Pass@1 accuracy and sustainability metrics, we evaluate whether models can reliably produce structured and directly executable code under a single-pass setting. This constraint reflects realistic usage in LLM-powered tooling (e.g., coding assistants), where outputs must be syntactically valid and conform to a predefined format to avoid additional post-processing, retries, or corrective prompting. All models are therefore evaluated using direct prompting via the DSPy \texttt{Predict} API to isolate each model’s intrinsic ability to follow output-format instructions without external scaffolding. We assess output correctness by first validating the Abstract Syntax Tree (AST) to ensure syntactic validity, and then checking for the presence of a callable Python function matching the expected interface. Failures are categorized as either parsing errors (invalid or malformed code) or test failures (syntactically valid code that does not satisfy functional requirements).

\begin{figure}[ht]
    \centering
    \includegraphics[width=1\linewidth]{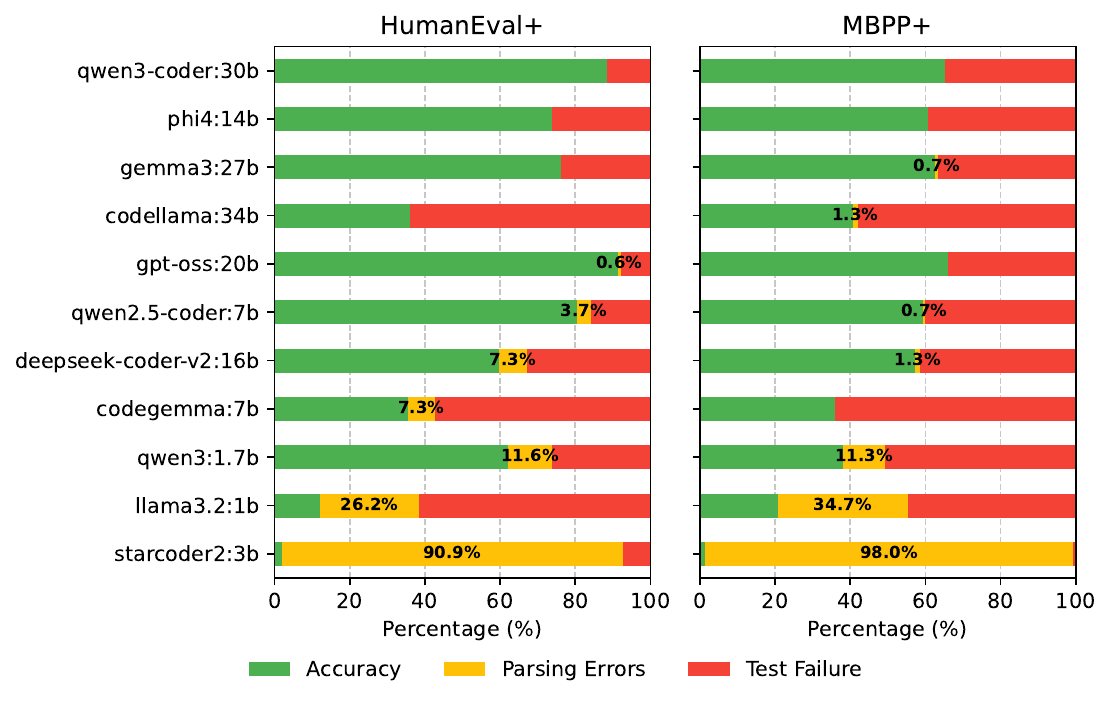}
    \caption{Comparison of parsing errors in different models.}
    \label{fig:model-parsing-error-comparison}
\end{figure}

\noindent\textbf{\textit{Observation.}} Figure~\ref{fig:model-parsing-error-comparison} shows that most small- and medium-sized SLMs produce well-formed, parsable outputs with low parsing error rates across both datasets. In particular, models such as Qwen, Phi, Gemma, and GPT-OSS exhibit near-zero parsing failures, indicating strong instruction-following capabilities despite their relatively small parameter counts.

In contrast, the smallest models—most notably StarCoder2:3B and LLama3.2:1B—exhibit substantially higher parsing error rates, with parsing failures exceeding 90\% on both benchmarks for StarCoder2:3B. These failures suggest that, at very small scales, models struggle to consistently satisfy strict output-format constraints, even when the task itself is relatively simple. From a systems perspective, such failures are consequential: malformed outputs necessitate re-prompting or additional validation and repair steps, which increase inference time, energy consumption, and overall emissions.

Overall, these observations justify our format-constrained, single pass evaluation setup for benchmarking SLMs and highlight output correctness as a critical but often overlooked dimension of model quality. Even when functional accuracy is comparable, models with poor formatting capability can impose hidden computational and environmental costs.

\section{Threats to Validity}

This study, while systematic, is subject to several threats to validity that should be considered when interpreting the results.

\noindent\textbf{Internal validity.} A potential confound arises from variability in model execution environments. Although all experiments were conducted under controlled conditions, background system processes and hardware-level fluctuations may have slightly affected power measurements and inference time. Another source of bias stems from the uneven performance of certain models (e.g., CodeLlama:34B), which may disproportionately influenced the aggregate statistics of the code-specialized category.

\noindent\textbf{External validity.} The generalizability of our findings is limited by the scope of models, tasks, and deployment settings. Our evaluation focuses on a subset of currently available open-source and general-purpose models and uses a specific benchmark configuration. Results may differ for other architectures, prompt engineering strategies, or deployment environments (e.g., cloud-based or multi-GPU settings). Furthermore, the selected dataset, while representative of real-world programming tasks, cannot fully capture the diversity and complexity of all software engineering problems.


\noindent \textbf{Construct Validity:} A potential threat to construct validity arises from our use of static Grid Carbon Intensity (GCI) values via CodeCarbon 3.1, which relies on historical grid averages. This approach overestimates emissions for rapidly decarbonizing grids such as Alberta’s, where the live 2026 GCI data would reduce absolute emissions by approximately 31.1\%. However, because Software Carbon Intensity (SCI) is modeled as a linear function of energy consumption with GCI as a constant scalar, and because the same CodeCarbon configuration was applied uniformly across all experiments, this limitation affects only the absolute emission magnitudes and not the relative comparisons, rankings, or trade-off conclusions between models and prompting strategies.

%% file: tables/carbon-accurate.tex
\begin{table}[!t]
\centering
\caption{Carbon per Correct Answer (CpCA) comparison across machines (using both datasets). Lower values indicate better sustainability.}
\label{tab:cpca_machine_comparison}
\resizebox{\columnwidth}{!}{
\begin{tabular}{lcccccc}
\toprule
& \multicolumn{3}{c}{\textbf{Machine 1}} & \multicolumn{3}{c}{\textbf{Machine 2}} \\
\cmidrule(lr){2-4} \cmidrule(lr){5-7}
\textbf{Strategy} & \textbf{CO$_2$} & \textbf{Pass@1} & \textbf{CpCA} & \textbf{CO$_2$} & \textbf{Pass@1} & \textbf{CpCA} \\
\midrule
Direct & 0.000212 & 61.84 & \textbf{0.000341} & 0.000054 & 60.45 & 0.000092 \\
CoT & 0.000230 & 63.94 & 0.000370 & 0.000050 & 64.11 & \textbf{0.000084} \\
PoT & 0.000586 & 59.26 & 0.001079 & 0.000132 & 59.36 & 0.000240 \\
LtoM & 0.001053 & 57.92 & 0.001773 & 0.000207 & 54.92 & 0.000383 \\
Self-Cons. & 0.001133 & 65.75 & 0.001752 & 0.000261 & 63.74 & 0.000444 \\
ReAct & 0.000882 & 59.62 & 0.001794 & 0.000230 & 58.06 & 0.000679 \\
\bottomrule
\end{tabular}
}
\end{table}

%% file: conclusion.tex
\section{Conclusions}

This paper presents the first comprehensive empirical analysis of how prompt engineering shapes the sustainability of code generation with small language models. By evaluating six prompting strategies across 11 open-source models with real-time environmental measurements on diverse hardware and in multiple regions, we demonstrate that code generation sustainability can be largely decoupled from accuracy. Simpler prompting strategies such as Chain-of-Thought can achieve near-optimal accuracy while reducing emissions by up to 80\% relative to more complex approaches like Self-Consistency. This reveals steep diminishing returns from heavyweight reasoning frameworks on small models. Overall, our findings suggest that practitioners should prioritize accuracy-per-watt rather than accuracy alone, as accuracy shows negligible correlation with environmental cost.

Our results also identify grid carbon intensity as the dominant driver of inference-time emissions, often outweighing hardware efficiency by an order of magnitude. This has important implications for geographically distributed open-source deployments. Model and prompting choices offer more immediate and controllable levers for sustainability than hardware upgrades. Yet ultimate progress remains constrained by regional energy infrastructure and the pace of clean energy transitions. Building on these insights, we outline several research directions, including standardized sustainability benchmarks, adaptive strategy selection based on real-time carbon signals, developer tooling that surfaces per-prompt emissions, broader hardware and lifecycle assessments, and incentive structures that favor specialized, efficient models over indiscriminate scaling. Together, these directions chart a path toward sustainable, accuracy-aware, and carbon-conscious code generation systems.

%% file: main.bbl
\begin{thebibliography}{10}

\bibitem{abdin2024phi4technicalreport}
M.~Abdin, J.~Aneja, H.~Behl, S.~Bubeck, R.~Eldan, S.~Gunasekar, M.~Harrison, R.~J. Hewett, M.~Javaheripi, P.~Kauffmann, J.~R. Lee, Y.~T. Lee, Y.~Li, W.~Liu, C.~C.~T. Mendes, A.~Nguyen, E.~Price, G.~de~Rosa, O.~Saarikivi, A.~Salim, S.~Shah, X.~Wang, R.~Ward, Y.~Wu, D.~Yu, C.~Zhang, and Y.~Zhang.
\newblock Phi-4 technical report, 2024.

\bibitem{agarwal2025gpt}
S.~Agarwal, L.~Ahmad, J.~Ai, S.~Altman, A.~Applebaum, E.~Arbus, R.~K. Arora, Y.~Bai, B.~Baker, H.~Bao, et~al.
\newblock gpt-oss-120b \& gpt-oss-20b model card.
\newblock {\em arXiv preprint arXiv:2508.10925}, 2025.

\bibitem{alibaba2026carbon}
{Alibaba Group}.
\newblock What is the carbon footprint difference between running local large language models and cloud-based apis per inference?
\newblock \url{https://www.alibaba.com/product-insights/}, Jan. 2026.
\newblock Alibaba Product Insights Blog.

\bibitem{ashraf2025energy}
H.~Ashraf, S.~M. Danish, A.~Leivadeas, Y.~Otoum, and Z.~Sattar.
\newblock Energy-aware code generation with llms: Benchmarking small vs. large language models for sustainable ai programming.
\newblock {\em arXiv preprint arXiv:2508.08332}, 2025.

\bibitem{austin2021program}
J.~Austin, A.~Odena, M.~Nye, M.~Bosma, H.~Michalewski, D.~Dohan, E.~Jiang, C.~Cai, M.~Terry, Q.~Le, et~al.
\newblock Program synthesis with large language models.
\newblock {\em arXiv preprint arXiv:2108.07732}, 2021.

\bibitem{chen2021evaluating}
M.~Chen.
\newblock Evaluating large language models trained on code.
\newblock {\em arXiv preprint arXiv:2107.03374}, 2021.

\bibitem{chen2022program}
W.~Chen, X.~Ma, X.~Wang, and W.~W. Cohen.
\newblock Program of thoughts prompting: Disentangling computation from reasoning for numerical reasoning tasks.
\newblock {\em arXiv preprint arXiv:2211.12588}, 2022.

\bibitem{columbia2023ai}
{Columbia Climate School, State of the Planet}.
\newblock Ai's growing carbon footprint, June 2023.
\newblock Accessed: 5 Dec 2025.

\bibitem{cruz2025prompt}
R.~Cruz, J.~Contreras, F.~Guerrero, E.~Rodriguez, C.~Valdez, and C.~Carrillo.
\newblock Prompt engineering and framework: implementation to increase code reliability based guideline for llms.
\newblock {\em arXiv preprint arXiv:2506.10989}, 2025.

\bibitem{dong2025survey}
Y.~Dong, X.~Jiang, J.~Qian, T.~Wang, K.~Zhang, Z.~Jin, and G.~Li.
\newblock A survey on code generation with llm-based agents.
\newblock {\em arXiv preprint arXiv:2508.00083}, 2025.

\bibitem{fu2025llmco2}
Z.~Fu, F.~Chen, S.~Zhou, H.~Li, and L.~Jiang.
\newblock Llmco2: Advancing accurate carbon footprint prediction for llm inferences.
\newblock {\em ACM SIGENERGY Energy Informatics Review}, 5(2):63--68, 2025.

\bibitem{giagnorio2025quantizing}
A.~Giagnorio, A.~Mastropaolo, S.~Afrin, M.~Di~Penta, and G.~Bavota.
\newblock Quantizing large language models for code generation: A differentiated replication.
\newblock {\em arXiv preprint arXiv:2503.07103}, 2025.

\bibitem{grattafiori2024llama3herdmodels}
A.~Grattafiori, A.~Dubey, A.~Jauhri, A.~Pandey, and et~al.
\newblock The llama 3 herd of models, 2024.

\bibitem{GreenSoftwareFoundation}
{Green Software Foundation}.
\newblock {Software Carbon Intensity (SCI) Specification}.
\newblock \url{https://sci.greensoftware.foundation/}, n.d.
\newblock Accessed: 2026-01-23.

\bibitem{hasan2025assessing}
M.~M. Hasan, M.~Waseem, K.-K. Kemell, J.~Rasku, J.~Ala-Rantala, and P.~Abrahamsson.
\newblock Assessing small language models for code generation: An empirical study with benchmarks.
\newblock {\em arXiv preprint arXiv:2507.03160}, 2025.

\bibitem{hoffmann2022training}
J.~Hoffmann, S.~Borgeaud, A.~Mensch, E.~Buchatskaya, T.~Cai, E.~Rutherford, D.~d.~L. Casas, L.~A. Hendricks, J.~Welbl, A.~Clark, et~al.
\newblock Training compute-optimal large language models.
\newblock {\em arXiv preprint arXiv:2203.15556}, 2022.

\bibitem{hui2024qwen2}
B.~Hui, J.~Yang, Z.~Cui, J.~Yang, D.~Liu, L.~Zhang, T.~Liu, J.~Zhang, B.~Yu, K.~Lu, et~al.
\newblock Qwen2. 5-coder technical report.
\newblock {\em arXiv preprint arXiv:2409.12186}, 2024.

\bibitem{husom2025sustainable}
E.~J. Husom, A.~Goknil, M.~Astekin, L.~K. Shar, A.~K{\~A}{\textyen}~sen, S.~Sen, B.~A. Mithassel, and A.~Soylu.
\newblock Sustainable llm inference for edge ai: Evaluating quantized llms for energy efficiency, output accuracy, and inference latency.
\newblock {\em ACM Transactions on Internet of Things}, 6(4):1--35, 2025.

\bibitem{khattab2023dspy}
O.~Khattab, A.~Singhvi, P.~Maheshwari, Z.~Zhang, K.~Santhanam, S.~Vardhamanan, S.~Haq, A.~Sharma, T.~T. Joshi, H.~Moazam, et~al.
\newblock Dspy: Compiling declarative language model calls into self-improving pipelines.
\newblock {\em arXiv preprint arXiv:2310.03714}, 2023.

\bibitem{lemos2025time}
F.~Lemos, V.~Alves, and F.~Ferraz.
\newblock Is it time to treat prompts as code? a multi-use case study for prompt optimization using dspy.
\newblock {\em arXiv preprint arXiv:2507.03620}, 2025.

\bibitem{li2024sprout}
B.~Li, Y.~Jiang, V.~Gadepally, and D.~Tiwari.
\newblock Sprout: Green generative ai with carbon-efficient llm inference.
\newblock In {\em Proceedings of the 2024 Conference on Empirical Methods in Natural Language Processing}, pages 21799--21813, 2024.

\bibitem{li2025structured}
J.~Li, G.~Li, Y.~Li, and Z.~Jin.
\newblock Structured chain-of-thought prompting for code generation.
\newblock {\em ACM Transactions on Software Engineering and Methodology}, 34(2):1--23, 2025.

\bibitem{liu2023improving}
C.~Liu, X.~Bao, H.~Zhang, N.~Zhang, H.~Hu, X.~Zhang, and M.~Yan.
\newblock Improving chatgpt prompt for code generation.
\newblock {\em arXiv preprint arXiv:2305.08360}, 2023.

\bibitem{liu2024exploring}
F.~Liu, Y.~Liu, L.~Shi, H.~Huang, R.~Wang, Z.~Yang, L.~Zhang, Z.~Li, and Y.~Ma.
\newblock Exploring and evaluating hallucinations in llm-powered code generation.
\newblock {\em arXiv preprint arXiv:2404.00971}, 2024.

\bibitem{lozhkov2024starcoder}
A.~Lozhkov, R.~Li, L.~B. Allal, F.~Cassano, J.~Lamy-Poirier, N.~Tazi, A.~Tang, D.~Pykhtar, J.~Liu, Y.~Wei, et~al.
\newblock Starcoder 2 and the stack v2: The next generation.
\newblock {\em arXiv preprint arXiv:2402.19173}, 2024.

\bibitem{meng2024empirical}
X.~Meng, Z.~Ma, P.~Gao, and C.~Peng.
\newblock An empirical study on llm-based agents for automated bug fixing.
\newblock {\em arXiv preprint arXiv:2411.10213}, 2024.

\bibitem{mu2024clarifygpt}
F.~Mu, L.~Shi, S.~Wang, Z.~Yu, B.~Zhang, C.~Wang, S.~Liu, and Q.~Wang.
\newblock Clarifygpt: A framework for enhancing llm-based code generation via requirements clarification.
\newblock {\em Proceedings of the ACM on Software Engineering}, 1(FSE):2332--2354, 2024.

\bibitem{ozcan2025quantifying}
M.~{\"O}zcan, P.~Wiesner, P.~Wei{\ss}, and O.~Kao.
\newblock Quantifying the energy consumption and carbon emissions of llm inference via simulations.
\newblock {\em arXiv preprint arXiv:2507.11417}, 2025.

\bibitem{patterson2022carbon}
D.~Patterson, J.~Gonzalez, U.~H{\"o}lzle, Q.~Le, C.~Liang, L.-M. Munguia, D.~Rothchild, D.~R. So, M.~Texier, and J.~Dean.
\newblock The carbon footprint of machine learning training will plateau, then shrink.
\newblock {\em Computer}, 55(7):18--28, 2022.

\bibitem{roh2025break}
J.~Roh, V.~Gandhi, S.~Anilkumar, and A.~Garg.
\newblock Break-the-chain: Reasoning failures in llms via adversarial prompting in code generation.
\newblock {\em arXiv preprint arXiv:2506.06971}, 2025.

\bibitem{roziere2023code}
B.~Roziere, J.~Gehring, F.~Gloeckle, S.~Sootla, I.~Gat, X.~E. Tan, Y.~Adi, J.~Liu, R.~Sauvestre, T.~Remez, et~al.
\newblock Code llama: Open foundation models for code.
\newblock {\em arXiv preprint arXiv:2308.12950}, 2023.

\bibitem{rubei2025prompt}
R.~Rubei, A.~Moussaid, C.~Di~Sipio, and D.~Di~Ruscio.
\newblock Prompt engineering and its implications on the energy consumption of large language models.
\newblock {\em arXiv preprint arXiv:2501.05899}, 2025.

\bibitem{su2024distilled}
C.-Y. Su and C.~McMillan.
\newblock Distilled gpt for source code summarization.
\newblock {\em Automated Software Engineering}, 31(1):22, 2024.

\bibitem{taherkhani2024epic}
H.~Taherkhani, M.~Sepindband, H.~V. Pham, S.~Wang, and H.~Hemmati.
\newblock Epic: Cost-effective search-based prompt engineering of llms for code generation.
\newblock {\em arXiv preprint arXiv:2408.11198}, 2024.

\bibitem{codegemmateam2024codegemmaopencodemodels}
C.~Team, H.~Zhao, J.~Hui, J.~Howland, N.~Nguyen, S.~Zuo, A.~Hu, C.~A. Choquette-Choo, J.~Shen, J.~Kelley, K.~Bansal, L.~Vilnis, M.~Wirth, P.~Michel, P.~Choy, P.~Joshi, R.~Kumar, S.~Hashmi, S.~Agrawal, Z.~Gong, J.~Fine, T.~Warkentin, A.~J. Hartman, B.~Ni, K.~Korevec, K.~Schaefer, and S.~Huffman.
\newblock Codegemma: Open code models based on gemma, 2024.

\bibitem{gemmateam2025gemma3technicalreport}
G.~Team, A.~Kamath, J.~Ferret, et~al.
\newblock Gemma 3 technical report, 2025.

\bibitem{vartziotis2024learn}
T.~Vartziotis, I.~Dellatolas, G.~Dasoulas, M.~Schmidt, F.~Schneider, T.~Hoffmann, S.~Kotsopoulos, and M.~Keckeisen.
\newblock Learn to code sustainably: An empirical study on llm-based green code generation.
\newblock {\em arXiv preprint arXiv:2403.03344}, 2024.

\bibitem{vartziotis2024carbon}
T.~Vartziotis, M.~Schmidt, G.~Dasoulas, I.~Dellatolas, S.~Attademo, V.~D. Le, A.~Wiechmann, T.~Hoffmann, M.~Keckeisen, and S.~Kotsopoulos.
\newblock Carbon footprint evaluation of code generation through llm as a service.
\newblock In {\em International Stuttgart Symposium}, pages 230--241. Springer, 2024.

\bibitem{wang2024selection}
C.-Y. Wang, A.~DaghighFarsoodeh, and H.~V. Pham.
\newblock Selection of prompt engineering techniques for code generation through predicting code complexity.
\newblock {\em arXiv preprint arXiv:2409.16416}, 2024.

\bibitem{wang2025comprehensive}
F.~Wang, Z.~Zhang, X.~Zhang, Z.~Wu, T.~Mo, Q.~Lu, W.~Wang, R.~Li, J.~Xu, X.~Tang, et~al.
\newblock A comprehensive survey of small language models in the era of large language models: Techniques, enhancements, applications, collaboration with llms, and trustworthiness.
\newblock {\em ACM Transactions on Intelligent Systems and Technology}, 16(6):1--87, 2025.

\bibitem{wang2024openhands}
X.~Wang, B.~Li, Y.~Song, F.~F. Xu, X.~Tang, M.~Zhuge, J.~Pan, Y.~Song, B.~Li, J.~Singh, et~al.
\newblock Openhands: An open platform for ai software developers as generalist agents.
\newblock {\em arXiv preprint arXiv:2407.16741}, 2024.

\bibitem{wang2022self}
X.~Wang, J.~Wei, D.~Schuurmans, Q.~Le, E.~Chi, S.~Narang, A.~Chowdhery, and D.~Zhou.
\newblock Self-consistency improves chain of thought reasoning in language models.
\newblock {\em arXiv preprint arXiv:2203.11171}, 2022.

\bibitem{wang2025codecontests+}
Z.~Wang, S.~Liu, Y.~Sun, H.~Li, and K.~Shen.
\newblock Codecontests+: High-quality test case generation for competitive programming.
\newblock {\em arXiv preprint arXiv:2506.05817}, 2025.

\bibitem{wei2022chain}
J.~Wei, X.~Wang, D.~Schuurmans, M.~Bosma, F.~Xia, E.~Chi, Q.~V. Le, D.~Zhou, et~al.
\newblock Chain-of-thought prompting elicits reasoning in large language models.
\newblock {\em Advances in neural information processing systems}, 35:24824--24837, 2022.

\bibitem{yang2025qwen3}
A.~Yang, A.~Li, B.~Yang, B.~Zhang, B.~Hui, B.~Zheng, B.~Yu, C.~Gao, C.~Huang, C.~Lv, et~al.
\newblock Qwen3 technical report.
\newblock {\em arXiv preprint arXiv:2505.09388}, 2025.

\bibitem{yao2022react}
S.~Yao, J.~Zhao, D.~Yu, N.~Du, I.~Shafran, K.~R. Narasimhan, and Y.~Cao.
\newblock React: Synergizing reasoning and acting in language models.
\newblock In {\em The eleventh international conference on learning representations}, 2022.

\bibitem{zhou2022least}
D.~Zhou, N.~Sch{\"a}rli, L.~Hou, J.~Wei, N.~Scales, X.~Wang, D.~Schuurmans, C.~Cui, O.~Bousquet, Q.~Le, et~al.
\newblock Least-to-most prompting enables complex reasoning in large language models.
\newblock {\em arXiv preprint arXiv:2205.10625}, 2022.

\bibitem{zhu2025towards}
B.~Zhu.
\newblock Towards principled training and serving of large language models.
\newblock 2025.

\bibitem{zhu2024deepseekcode}
Q.~Zhu, D.~Guo, Z.~Shao, D.~Yang, P.~Wang, R.~Xu, Y.~Wu, Y.~Li, H.~Gao, S.~Ma, et~al.
\newblock Deepseek-coder-v2: Breaking the barrier of closed-source models in code intelligence.
\newblock {\em arXiv preprint arXiv:2406.11931}, 2024.

\end{thebibliography}
